\documentclass[aps,prb,twocolumn,showpacs, superscriptaddress]{revtex4}

\usepackage{amsmath,amssymb}
\usepackage{graphicx}
\usepackage{bm}
\usepackage{epstopdf}
\usepackage{color}

\newcommand{\slrr}      {$T_1^{-1}$}
\newcommand{\y}        {$^{89}$Y}
\newcommand{\bayiro}	{Ba$\mathrm{_{2}}$YIrO$\mathrm{_{6}}$}
\newcommand{\sryiro}	{Sr$\mathrm{_{2}}$YIrO$\mathrm{_{6}}$}

\begin{document}

\title{Diluted paramagnetic impurities in nonmagnetic \bayiro\ }

\author{F. Hammerath}
\email[Corresponding author: ]{f.hammerath@ifw-dresden.de}
\affiliation{Institute of Solid State and Materials Physics, TU-Dresden, 01062 Dresden, Germany}
\affiliation{IFW-Dresden, Institute for Solid State Research, Helmholtzstra{\ss}e 20, 01069 Dresden, Germany}
\author{R. Sarkar}
\affiliation{Institute of Solid State and Materials Physics, TU-Dresden, 01062 Dresden, Germany}
\author{S. Kamusella}
\affiliation{Institute of Solid State and Materials Physics, TU-Dresden, 01062 Dresden, Germany}
\author{C. Baines}
\affiliation{Laboratory for Muon Spin Spectroscopy, Paul Scherrer Institute, CH-5232 Villigen, PSI, Switzerland}
\author{H.-H. Klauss}
\affiliation{Institute of Solid State and Materials Physics, TU-Dresden, 01062 Dresden, Germany}
\author{T. Dey}
\affiliation{IFW-Dresden, Institute for Solid State Research, Helmholtzstra{\ss}e 20, 01069 Dresden, Germany}
\author{A. Maljuk}
\affiliation{IFW-Dresden, Institute for Solid State Research, Helmholtzstra{\ss}e 20, 01069 Dresden, Germany}
\author{S. Ga{\ss}}
\affiliation{IFW-Dresden, Institute for Solid State Research, Helmholtzstra{\ss}e 20, 01069 Dresden, Germany}
\author{A.U.B. Wolter}
\affiliation{IFW-Dresden, Institute for Solid State Research, Helmholtzstra{\ss}e 20, 01069 Dresden, Germany}
\author{H.-J. Grafe}
\affiliation{IFW-Dresden, Institute for Solid State Research, Helmholtzstra{\ss}e 20, 01069 Dresden, Germany}
\author{S. Wurmehl}
\affiliation{Institute of Solid State and Materials Physics, TU-Dresden, 01062 Dresden, Germany}
\affiliation{IFW-Dresden, Institute for Solid State Research, Helmholtzstra{\ss}e 20, 01069 Dresden, Germany}
\author{B. B\"{u}chner}
\affiliation{Institute of Solid State and Materials Physics, TU-Dresden, 01062 Dresden, Germany}
\affiliation{IFW-Dresden, Institute for Solid State Research, Helmholtzstra{\ss}e 20, 01069 Dresden, Germany}
\date{\today}

\begin{abstract}

The cubic double perovskite \bayiro\ has been investigated by the local probe techniques nuclear magnetic resonance (\y\ NMR) and muon spin rotation ($\mu$SR). Both methods confirm the absence of magnetic long-range order in this compound but find evidence for diluted localized paramagnetic moments.
NMR spin-lattice relaxation rate \slrr\ measurements suggest a slowing down of localized spin moments at low temperatures. An increase of the $\mu$SR spin-lattice relaxation rate $\lambda$ confirms the presence of weak magnetism in \bayiro. However, these findings cannot be explained by the recently suggested excitonic type of magnetism. Instead, they point towards tiny amounts of localized paramagnetic spin centers leading to this magnetic response on the background of a simple nonmagnetic ground state of the 5$d^4$ ($J=0)$ electronic configuration of Ir$^{5+}$.

\end{abstract}


\maketitle

\section{Introduction}

In recent years the 5$d$ transition metal oxides have gained a lot of interest since the enhanced spin-orbit coupling (SOC) in these systems adds an interesting new energy scale to the rich physics of competition between crystal field (CF) and Coulomb interactions (U). This may lead to unexpected exotic magnetic phases and novel types of local-moment frustration.\cite{Jackeli2009, Chaloupka2010}

In most cases, systems with an odd number of electrons in the $d$ shell (Ir$^{4+}$, 5$d^5$) have been studied, such as {\it A}$_2$IrO$_4$ ({\it A} = Sr, Ba),\cite{Kim2008PRL, Jackeli2009, Okabe2011, Boseggia2013} {\it A}$_2$IrO$_3$ ({\it A} = Na, Li)\cite{Shitade2009, Chaloupka2010, Singh2010, Singh2012, Choi2012, Gretarsson2013} and BaIrO$_3$.\cite{Cao2000, Brooks2005}
In the case of a 5$d^4$ (Ir$^{5+}$) electronic configuration, the strong SOC of the $t_{2g}$ state with an integer low-spin state $S$ leads to a splitting into a nonmagnetic $J=0$ ground state and magnetic $J=1$ (and $J=2$) excited states.\cite{Khaliullin2013} Magnetism can form only via a Van-Vleck-like singlet - triplet excitation over the SOC gap. Since this singlet - triplet gap is of similar size as the spin exchange energy scale ($\sim 50 - 100$\,meV), unusual magnetic states can be expected.\cite{Meetei2015}
The double perovskites {\it A}$_2${\it BB}'O$_6$ (with {\it A} an alkaline earth element and {\it B} and {\it B}' two different transition metal ions) were suggested as potential candidates for such unusual magnetic states.\cite{Meetei2015}

From the experimental side, very different interpretations concerning these double perovskites have been raised.
Early magnetization and heat capacity measurements on single crystals of Sr$_2$YIrO$_6$ and Ba$_{2-x}$Sr$_x$YIrO$_6$ ($x=$0, 0.74) revealed an antiferromagnetically long-range ordered ground state below 1.3\,K (1.6\,K respectively),\cite{Cao2014, Terizc2016} in contrast to the expected $J =0$ nonmagnetic ground state.\cite{Cao2014} In the case of \sryiro\ this was assigned to result from the high distortion of the IrO$_6$ octahedra in the monoclinic structure of this compound.\cite{Cao2014} De facto, lattice degrees of freedom can play a significant role in determining the ground state and the physical properties of a material. Slight alterations of the lattice may lead to strong effects in the magnetic properties, especially due to the extended 5$d$ orbitals, as found e.g. in Sr- and Ru-doped BaIrO$_3$.\cite{Cao2004, Yuan2016} However, this explanation breaks down in the case of the cubic analogue \bayiro\ with regular IrO$_6$ octahedra. Terizc {\it et al.}, thus, interpret the observed long-range magnetic order in Ba$_{2-x}$Sr$_x$YIrO$_6$ as a result of band structure and/or electron-electron interaction effects competing with the SOC.\cite{Terizc2016}
In contrast, studies on polycrystals of the series Ba$_{2-x}$Sr$_x$YIrO$_6$ reported the absence of any long-range order and no significant change in the magnetic properties upon increasing the structural disorder from cubic \bayiro\ with regular IrO$_6$ octahedra to monoclinic \sryiro\ with highly distorted IrO$_6$ octahedra, thus demonstrating that the CF splitting in \sryiro\ is not enough to compete with the strong spin-orbit coupling.\cite{Ranjbar2015, Phelan2016}   
A recent detailed single crystal x-ray diffraction analysis of \sryiro\ even questioned the formerly reported monoclinic structure\cite{Wakeshima1999, Cao2014, Ranjbar2015} but reported a cubic unit cell with regular IrO$_6$ octahedra.\cite{Corredor2017} Furthermore, no evidence for magnetic long-range order has been found down to 430\,mK and the low temperature anomaly in the specific heat has been identified as a Schottky anomaly caused by tiny amounts ($<1$\,\%) of paramagnetic impurities.\cite{Corredor2017}  
Also for single crystals of \bayiro\ long-range magnetic order has not been observed down to 0.4\,K and the effective magnetic moment of 0.44\,$\mu _B$/Ir (extracted from a Curie-Wei\ss\ (CW) fit to the magnetic susceptibility), could be accounted for by $\sim$2\,\% of $J=1/2$ impurities.\cite{Dey2016}  These studies thus corroborate the dominance of the SOC leading to a nonmagnetic $J=0$ ground state, whose weak magnetic response stems from paramagnetic impurities, such as tiny amounts of Ir$^{4+}$ or Ir$^{6+}$ ions, created by chemical disorder (Ir$^{4+}$ for Y$^{3+}$ substitution) and/or off-stoichiometry, respectively.\cite{Dey2016}

We investigated this weak magnetism by means of nuclear magnetic resonance and muon spin rotation measurements on single crystals and polycrystals of \bayiro. Taking advantage of the local probe character of these two methods in contrast to bulk characterization techniques, we can directly reveal the nonmagnetic $J$=0 ground state of \bayiro, and show, that the observed weak magnetism stems from small amounts of diluted paramagnetic spin centers on top of this nonmagnetic ground state. The origin of these paramagnetic spin centers may in principle be either disorder, off-stoichiometry, oxygen deficiency or excess, although some scenarios seem to be more likely than others.

\section{Sample preparation, characterization and experimental details}
\label{sampleprep}

High purity starting materials BaCO$_3$ (Alpha Aesar 99.997\,\%), IrO$_2$ (Alpha Aesar 99.99\,\%) and Y$_2$O$_3$ (Alpha Aesar 99.999\,\%) were used to grow single crystals of \bayiro\ in a flux of ultradry BaCl$_2$ (Alpha Aesar 99.5\,\%) as described in detail in Ref.~\onlinecite{Dey2016}. Single crystals of typical dimensions 0.3 x 0.3 x 0.3 mm$^3$ were obtained.\cite{Dey2016}
Details on their characterization and magnetic properties are found in Ref.~\onlinecite{Dey2016}. 
A polycrystalline sample of \bayiro\ has been prepared by solid state reaction method. Stoichiometric amounts of high-purity BaCO$_3$, Y$_2$O$_3$, and Ir metal powder were mixed thoroughly, pressed into pellets and calcined at 800\,$^\circ$C for 12h. Subsequently, the mixture was heated at 1200\,$^\circ$C for 80h with several intermediate grinding and pelletizing. From the point of view of standard bulk characterization techniques (XRD and magnetization) it appeared to be less disordered than the single crystals. 

For \y\ nuclear magnetic resonance (NMR) measurements plenty of single crystals were crushed to a powder. 
The crushed single crystals were measured in static magnetic fields of 7 and 15\,T, while the polycrystalline sample was measured in 7\,T. \y\ possesses a nuclear spin $I =1/2$, thus no quadrupolar effects influenced the NMR spectra and relaxation rate measurements. 
NMR spectra were measured by using a normal Hahn spin-echo sequence and Fourier transforming the echo. The repetition rate between subsequent pulse sequences was chosen to be suffienctly long to prevent spin-lattice relaxation rate effects. The spin lattice relaxation rate \slrr\ was measured by using the saturation recovery method and the recovery of the nuclear magnetization $M_z(t)$ was fitted to:
\begin{equation}
M_z(t) = M_0[1-fe^{-(t/T_1)^\beta}] \label{eq:T1}
\end{equation}
with the nuclear saturation magnetization $M_0$, $f=1$ for an ideal saturation recovery (otherwise $f<1$) and a stretched exponent $\beta$. $0<\beta < 1$ refers to a probability distribution $P$ of individual spin-lattice relaxation rates $T_{1,i}^{-1}$, with which the stretched exponential function can be expressed as: $e^{-(t/T_1)^\beta} = \int_0^\infty P(s,\beta)e^{-st/T_1}ds$, where $s=T_1/T_{1,i}$ and $\int_0^\infty P(s,\beta)ds = 1$.\cite{Johnston2005, Johnston2006, Shiroka2011} In this case, \slrr\ denotes a characteristic relaxation rate of the system. Please note, that this characteristic relaxation rate differs from the average relaxation rate or its inverse. Instead, for $1/3 < \beta < 1$, \slrr\  can be understood as the value for which it is equally likely for $T_{1,i}^{-1}$ to be smaller or greater than \slrr. For $\beta = 1$, the probability distribution amounts to the Dirac function at $s = 1$ and yields $T_{1,i}^{-1}$ = \slrr. With decreasing $\beta$, the probability distribution $P(s,\beta)$ gets broader and more and more asymmetric and its maximum shifts to slower rates $s<1$. Former works used a log-normal distribution to mimic the probability distribution of the stretched exponential.\cite{Mitrovic2008, Dioguardi2015} They showed that the extracted $T_{1,lognorm}^{-1}$ are very similar to the characteristic \slrr\ as extracted by a stretched exponential recovery law and do not depend on the exact form of the distribution function. We therefore decided to use the stretched exponential fitting function for this work. For a more detailed analysis of the probability distribution the interested reader is referred to Ref. \onlinecite{Johnston2006}.

$\mu$SR measurements on the same crushed single crystals of \bayiro\ have been carried out at the $\pi$M3 beam line at the GPS and LTF spectrometers of the Swiss Muon Source at the Paul Scherrer Institute (PSI) in Villigen. Positively spin-polarized muons were implanted into the sample and the time evolution of the muon spin polarization, $P(t)$, was monitored by detecting the asymmetric spatial distribution of positrons emitted from the muon decay.\cite{Schenck1985, Yaouanc2011} The measurements have been performed in the temperature range between 300\,mK and 200\,K in zero magnetic field (ZF) and in longitudinal applied magnetic fields with respect to the initial muon spin polarization up to 1000\,G (LF). To improve the thermal contact in the LTF machine, the samples were glued on an Ag plate giving rise to a time and temperature independent background signal due to muons that stopped in the Ag plate. The $\mu$SR time spectra were analyzed using the free software package MUSRFIT.

\section{Experimental Results and Discussion}

\subsection{Nuclear Magnetic Resonance}

\begin{figure}[t]
\centering
\includegraphics[width=\linewidth]{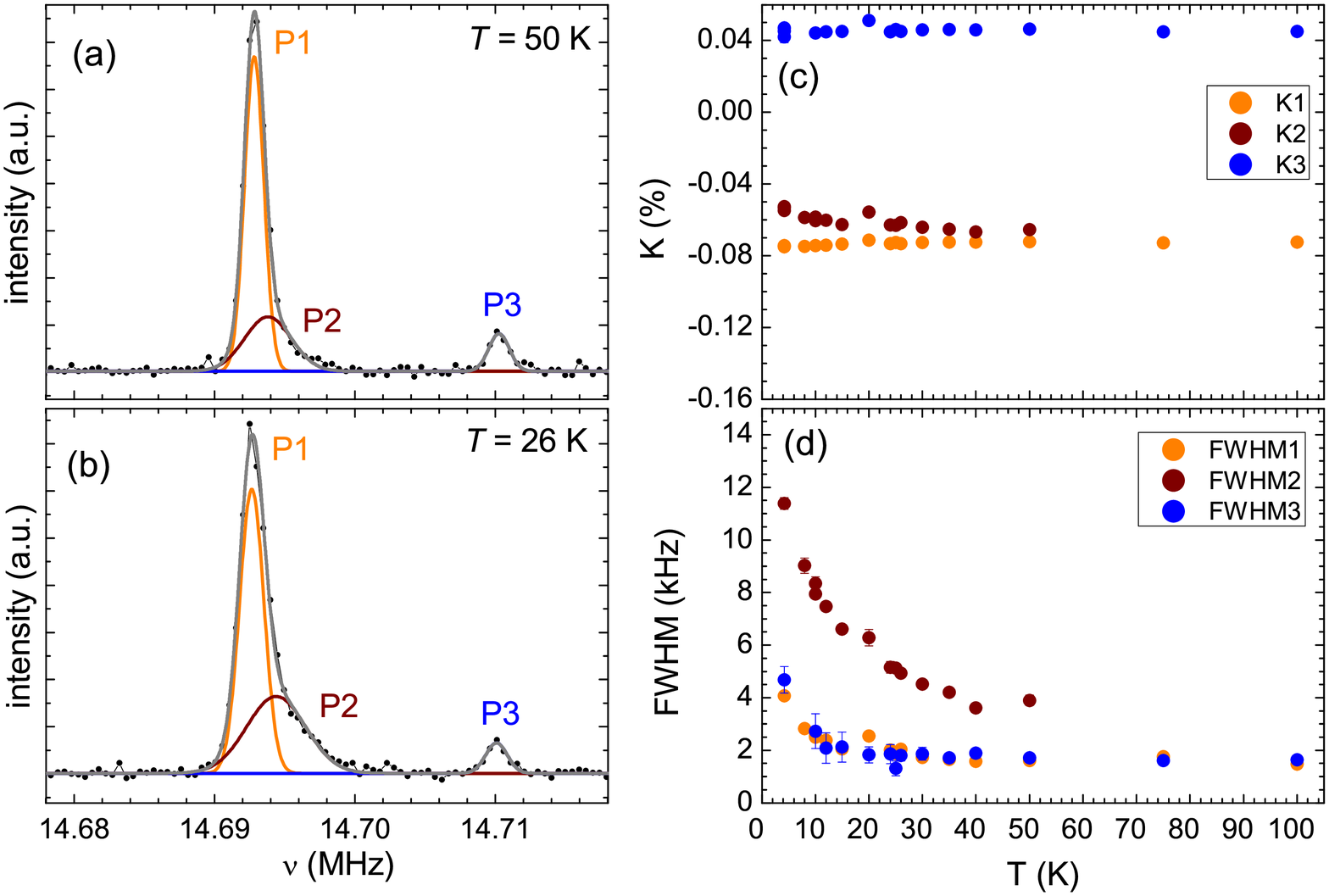}
\caption{\label{fig:NMRspectra} (Color online) Left column (a), (b): Selected \y\ NMR spectra of crushed single crystals of \bayiro, measured in a field of 7.0493\,T. Lines are fits with three Gaussians to the data, resulting in an anisotropic main line [peaks P1 (orange) and P2 (dark red)], as well as an additional peak at higher frequencies (P3, blue). 
Right column: (c) Knight shift and (d) FWHM as function of temperature as extracted from the Gaussian fits. Note that due to the reduced signal-to-noise ratio at high temperatures, peak P2 could not be resolved for 75\, and 100\,K.}
\end{figure}

NMR measurements have been performed in the temperature range from 4.2\,K up to 100\,K. Figs.~\ref{fig:NMRspectra}(a) and (b) show representative spectra measured in 7.0493\,T. 
The main resonance line stemming from Y nuclei within the intrinsic \bayiro\ phase is peaked around 14.692\,MHz. It is asymmetric and can be described by a fit with two Gaussian lines, revealing the main peak (P1) and a high-frequency shoulder (P2). Additionally, a smaller peak is found at slightly higher frequencies (P3), with an average spectral weight of roughly 5\,\%. The nuclear spin-lattice relaxation rate of this small high-frequency peak is not accurately measurable. It is of the order of several thousands of seconds. Its observed temperature independent Knight shift K3 $\simeq~0.04\%$ [see Fig.~\ref{fig:NMRspectra}(c)] agrees well with the reported chemical shift of Y$_2$O$_3$ (roughly 300 ppm).\cite{Harazono1997} Furthermore, traces of unreacted Y$_2$O$_3$ of the same order of magnitude (about 2\,\%) have also been found in the powder XRD characterization of these samples.\cite{Dey2016} Hence, we attribute P3 to Y nuclei within a tiny nonmagnetic Y$_2$O$_3$ impurity phase.

Fig.~\ref{fig:NMRspectra}(c) shows the NMR Knight shift $K$ of all three peaks, extracted from their measured resonance frequencies $\nu$ as: 
\begin{equation}
\omega =  2\pi\nu = \gamma H_0 (1+K) \, ,
\end{equation}
where $\gamma/2\pi = 2.0858$\,Mhz/T is the gyromagnetic ratio of the \y\ nuclei and $\mu_0H_0$ = 7.0493\,T is the magnetic field.
The Knight shift results from the hyperfine interaction between the spins of the electrons and the nuclear spins. In an applied magnetic field, polarized electrons create an additional hyperfine field at the nuclear sites, leading to a shift of the resonance line with respect to the unshifted Larmor frequency $\omega_L = \gamma H_0$. This shift $K$ comprises the temperature-independent contributions: $K_{orb}$ (orbital shift) and $K_{dia}$ (diamagnetic shift), as well as a (in most cases temperature-dependent) spin part $K_s$, which is a measure of the intrinsic local spin susceptibility $\chi_s(q=0, \omega =0)$:
\begin{equation} 
K_s = A_{hf} \chi_s(q=0, \omega =0) \, ,
\label{eq:hyperfine}
\end{equation}
with the hyperfine coupling constant $A_{hf}$. $K_s$ itself can consist of dipolar contributions, Fermi-contact contributions (only in the case of unpaired $s$-electrons), and transferred Fermi-contact contributions via the polarization of the orbitals (also called core polarization).

\begin{figure}[b]
\centering
\includegraphics[width=0.8\linewidth]{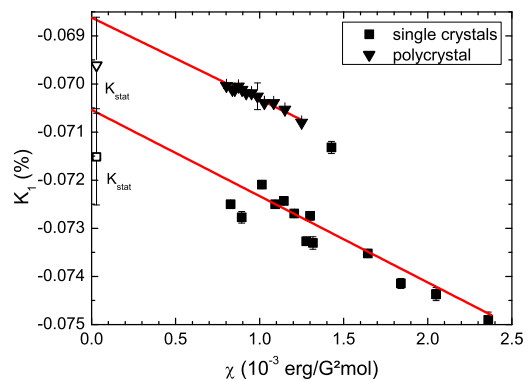}
\caption{\label{fig:hyperfine} (Color online) Knight shift $K_1$ of the NMR resonance line of the main \bayiro\ phase as a function of the macroscopic susceptibility $\chi$ with temperature as implicit parameter for the single crystals (full squares) and the polycrystal (full triangels). Lines are fits according to Eq.~\ref{eq:hyperfine}, including a constant offset. The open symbols denote the deduced range for the temperature independent shift $K_{stat} = K_{orb} + K_{dia}$.}
\end{figure}

The nearly temperature independent Knight shifts of the main \bayiro\ phase, $K1$ and $K2$, and their low absolute values indicate that the local spin susceptibility is very small and only weakly temperature dependent, corroborating the theoretically proposed nonmagnetic $J=0$ ground state. This stands in contrast to the measured bulk susceptibility,  which cannot differentiate between the intrinsic spin susceptibility and contributions from even tiny amounts of paramagnetic impurities, leading to a Curie-Wei{\ss}-like upturn at low temperatures.\cite{Dey2016} In the high temperature range, where $\chi$ is still rather flat, the hyperfine coupling constant can be deduced according to Eq.~\ref{eq:hyperfine} and including a constant term for the static part $K_{stat}$.\cite{Johnston2010} Fig.~\ref{fig:hyperfine} shows $K_1$ as a function of the macroscopic susceptibility $\chi$  with temperature as implicit parameter (for T $>$ 10\,K). The deduced hyperfine coupling constant amounts to $A_{hf} = -0.1(2)$\,kOe/$\mu_B$. This small negative value is the sum of (direction-dependent negative and positive) dipolar and (negative) core polarization hyperfine interactions, which cannot be separated from each other. By extrapolating to $\chi = 0$ and considering the maximal possible values of $K_s$, we can also deduce the temperature independent part of the Knight shift, $K_{stat}$. This part comprises the orbital shift $K_{orb}$ from unfilled electronic shells and the diamagnetic shift $K_{dia}$ stemming from closed inner electronic shells. $K_{stat}$ amounts to (-0.071 $\pm$ 0.002)\,\%. Hence, it is the dominant contribution to the measured shift, while the spin part $K_s$ is vanishingly small. These results together with the weak temperature dependence strongly hint towards a nonmagnetic ground state of \bayiro.  

Solely the Knight shift of the shoulder peak P2 shows a slight enhancement towards low temperature, suggesting the presence of diluted paramagnetic impurities.
We thus assume that the main peak P1 stems from nuclei in the nonmagnetic $J = 0$ \bayiro\ matrix, while peak P2 might stem from nuclei which sense the previously suggested Y-Ir disorder, leading to a few percent of paramagnetic J=1/2 impurity centers.\cite{Dey2016} 
In fact, an energy dispersive x-ray (EDX) analysis with a scanning electron microscope (SEM) suggested a small excess of Y in the crystal. Thus, a certain off-stoichiometry, i.e. Ba$_2$Y$_{1+d}$Ir$_{1-d}$O$_6$, and/or site disorder between Y and Ir in the crystals cannot be ruled out, although the refinement of the powder XRD pattern could not reveal such effects.\cite{Dey2016}
Note that in principal also a slight oxygen deficiency could lead to paramagnetic Ir$^{4+}$ ions. However, previous magnetization measurements comparing as-grown crystals and crystals annealed in oxygen pressure did not show any difference,\cite{Dey2016} rendering this possibility rather unlikely. Since Ir$^{6+}$ is also magnetic ($J$ = 3/2), one may assign the paramagnetic impurities to stem from Ir$^{6+}$ caused by oxygen excess, however experimental evidence for this scenario is lacking so far and former growth studies showed, that  Ir$^{6+}$ can only be stabilized under high oxygen pressure and high temperature.\cite{Demazeau1993}  

Note that in light of the XRD results and the magnetization measurements,\cite{Dey2016} the spectral weight distribution between P1 and P2 of 70\,\% : 30\,\% seems surprisingly large. However, if we assume that, due to the extended 5$d$ orbitals, also nearest neighbors (nn) [and potentially next-nearest neighbors (nnn)] sense the disorder caused by some tiny Y-Ir mismatch, the effects seen by NMR will be larger than the actual disorder itself. In fact, by considering the coordination number six at the Y site, P1 can be attributed to an Y - (O-Ir)$_6$ nn environment, while P2 can be assigned to an Y - (O-Ir)$_5$-(O-Y) nn environment and the spectral weight distribution between P1 and P2 translates to a Y-Ir mismatch of $\sim$5\%.\cite{Aharen2009,Aharen2010}

Fig.~\ref{fig:NMRspectra}(d) shows the full width at half maximum (FWHM) of all three peaks. At high temperatures, the resonance lines are very narrow, pointing towards a nonmagnetic environment of the \y\ nuclei, since any magnetic environment would lead to slightly different hyperfine fields and thus slightly different $K$, resulting in a broadening of the NMR lines. 
The linewidth of the shoulder peak P2 is always larger than the ones of P1 and P3, suggesting the more magnetic environment of the \y\ nuclei at this site, likely stemming from Y-Ir disorder. 
A small line broadening is observed at lower temperatures, pointing towards a slowing down of magnetic fluctuations. 
The absolute values of the linewidth at low temperature [11.4\,kHz at 4.2\,K (P2)] can be well reproduced by considering the deduced hyperfine coupling constant and the magnetic moment as obtained from a Curie-Wei\ss\ fit to the macroscopic susceptibility, which amounts to $\mu_{eff}$ = 0.44\,$\mu_B$/Ir atom.\cite{Dey2016} The internal field $H_{int} = A_{hf}\mu_{eff}$ = 44\,Oe leads to a line broadening due to the effective magnetic moment of $\Delta \nu = \gamma H_{int}$ = 9.2\,kHz. Taken together with the linewidth in the high temperature limit, which is based on other effects, this can account for the entire linewidth of 11.4\,kHz at 4.2\,K. Hence, no other line broadening effects have to be considered. The observed line broadening is solely a result of the tiny amount of paramagnetic impurities.

\begin{figure}[t]
\centering
\includegraphics[width=0.8\linewidth]{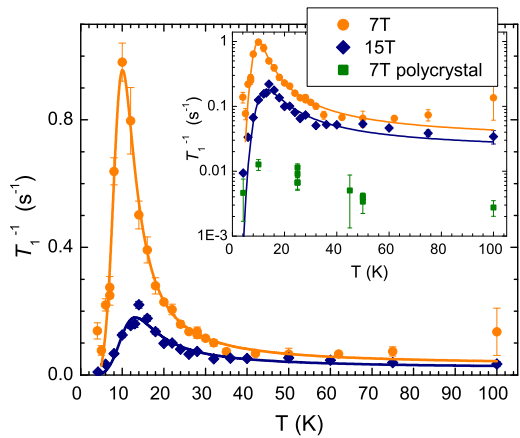}
\caption{\label{fig:NMRT1} (Color online) \y\ NMR spin-lattice relaxation rate \slrr\ of \bayiro, measured in magnetic fields of 7.0493\,T (orange dots) and 15\,T (dark blue diamonds). Lines are fits to the data according to Eq.~\ref{eq:BPP}. The inset shows the same data and fits with a logarithmic ordinate. Additionally, \slrr\ of a polycrystalline \bayiro\ sample in 7.0493\,T is shown in the inset (green squares).}
\end{figure}

Measurements of the NMR spin-lattice relaxation rate $T_1$ have been performed on the main peak in magnetic fields of 7 and 15\,T (see Fig.~\ref{fig:NMRT1}). Since P1 and P2 are very close to each other in the spectrum, $T_1^{-1}$ can only be measured for both of them at the same time. This results in a stretched exponential recovery curve (see Eq.~\ref{eq:T1}) with a basically temperature-independent stretching exponent 
$0.4 < \beta < 0.6$, which points towards a rather broad, temperature-independent probability distribution of individual relaxation rates $T_{1,i}^{-1}$. The characteristic spin-lattice relaxation rate of this distribution, \slrr, is temperature independent from 100\,K down to roughly 40~\,K. The absolute values of $T_1 \simeq 15$\,s (20\,s) 
in 7\,T (15\,T) in this temperature range indicate the absence of any type of strong magnetic correlations, which would lead to a much faster relaxation. 

\begin{table}[b]
    \center
    \setlength{\tabcolsep}{5pt}
    \caption{Temperature of the maximum of the BPP peak $T_{max}$ and BPP fitting parameters of Eq.~\ref{eq:BPP}.}
    \begin{tabular}{c c c c c}
     \hline
     \hline
      & $T_{max}$ (K) & $E_a$ (K) & $\tau_0$ ($10^{-10}$) s & $h_{\perp}$ (Gauss)\\
     \hline
     7\,T& 10 & $(42.4 \pm 1.3)$ & $(1.6 \pm 0.2)$ & $(10.3 \pm 0.2)$\\
     15\,T & 14 & $(38.4 \pm 2.7)$ & $(2.7 \pm 0.4)$ & $(6.4 \pm 0.2)$\\
     \hline
     \hline
   \end{tabular}
   \label{tab:tab}
\end{table}

However, upon further lowering the temperature, $T_1$ gets faster and a pronounced peak appears in \slrr\ for both fields.\cite{remark} Such a peak points towards a progressive slowing down of spin fluctuations, which can be described in the Bloembergen-Purcell-Pound (BPP) model.\cite{BloembergenNature1947, Bloembergen1948} Within this model, the spin-lattice relaxation rate \slrr\ is expressed as a function of the local fluctuating magnetic field $h_{\perp}(t)$ perpendicular to the applied magnetic field and its characteristic autocorrelation time $\tau_c$:\cite{BloembergenNature1947, Bloembergen1948}
\begin{equation}
T_{1, BPP}^{-1}(T) = \gamma^2 h_{\perp}^2 \frac{\tau_c(T)}{1+\tau_c^2(T)\omega_L^2} \, .
\label{eq:BPP}
\end{equation} 
This leads to a peak in \slrr\ at the temperature where the correlation time of the spin fluctuations $\tau_c$ equals the inverse Larmor frequency $\omega_L$. For disordered systems, the temperature dependence $\tau_c(T)$ can be well described by an activated behavior:\cite{Suh2000, Curro2000, Curro2009, Hammerath2013, Hammerath2015}
$\tau_c(T) = \tau_0 \exp(E_a/k_BT)$,
with the activation energy $E_a$ of the magnetic fluctuations and the correlation time at infinite temperature $\tau_0$. Thus, the peak in \slrr\ should shift to  higher temperature upon increasing the external magnetic field, which is indeed observed experimentally (see Fig.~\ref{fig:NMRT1} and results in Table~\ref{tab:tab}).

\begin{figure}[t]
\centering
\includegraphics[width=0.9\linewidth]{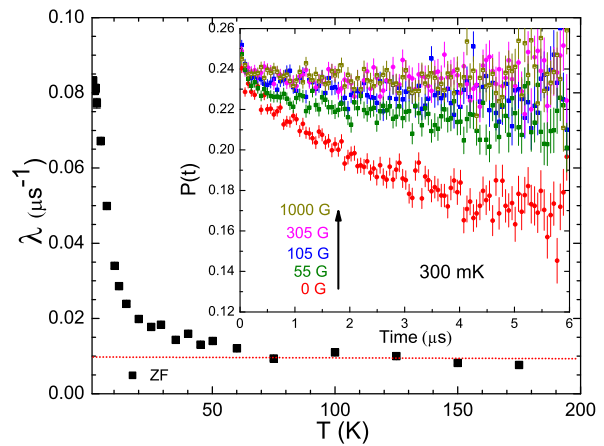}
\caption{\label{fig:ZFT1} (Color online) ZF $\mu$SR  spin-lattice relaxation rate $\lambda$ as a function of temperature. The inset shows the decoupling experiments at 300\,mK. }
\end{figure}

Fits according to Eq.~\ref{eq:BPP} (solid lines in Fig.~\ref{fig:NMRT1} and Table~\ref{tab:tab}) can well describe the data and yield an activation energy of the slowing down of magnetic fluctuations of about $E_a = (40 \pm 4)$\,K, which is in good agreement with the onset temperatures of the increase of the NMR FWHM [see. Fig.~\ref{fig:NMRspectra}(d)] and the $\mu$SR spin-lattice relaxation rate $\lambda$ (see Fig.~\ref{fig:ZFT1}). 
The fluctuating field $h_{\perp}$ amounts to around $(8 \pm 2)$\,Gauss, which is fairly small. 
It compares well to the internal field as extracted from the linewidth of P2 by considering that $h_{\perp}$ has been deduced from spin-lattice relaxation rate measurements on P1 and P2, while $H_{int}$ has been extracted from the linewidth of P2 only. The extracted activation energy of the slowing down of spin fluctuations, $E_a = (40 \pm 4)$\,K is too small to account for singlet - triplet excitations over the SOC gap, which is of the order of 370\,meV.\cite{Kusch2017, Nag2017} 
 Since the BPP peak is seen at the main NMR resonance lines P1 and P2 of the \bayiro\ phase, we suppose that these fluctuations stem from homogeneously distributed paramagnetic spin centers in the main phase. As already cited, these spin centers may arise due to slight off-stoichiometry and chemical disorder, leading to Ir$^{4+}$ and Ir$^{6+}$ ions with $J=1/2$ and $J=3/2$, respectively. Recent electron spin resonance (ESR) investigations come to the same conclusion and give a detailed analysis of these spin centers.\cite{Fuchs2017}
Certainly, the system is far from being magnetically long-range ordered, with the stoichiometric \bayiro\ phase being in a $J=0$ nonmagnetic ground state as expected from the strong SOC. 

\begin{figure}[t]
\centering
\includegraphics[width=0.8\linewidth]{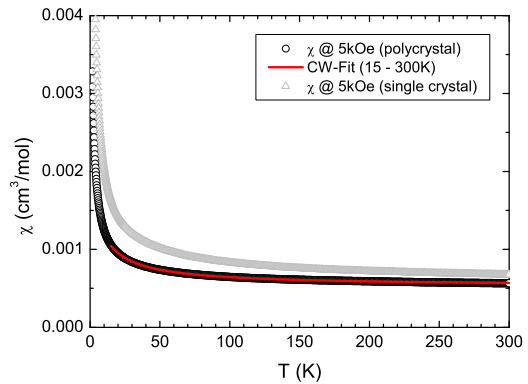}
\caption{\label{fig:polysusc} (Color online) Comparison of the temperature dependent magnetic susceptibility of a polycrystalline \bayiro\ sample (black dots) with the one of the single crystal (grey triangles) in an external magnetic field of 5\,kOe. The red line shows the CW-Fit to the polycrystal data in the range from 15 - 300\,K.} \end{figure}

For comparison, we also measured a polycrystalline sample of \bayiro, which from the point of view of the bulk characterization appears to be less disordered than the single crystals. Its magnetic susceptibility is smaller than the one of the single crystal (see Fig.~\ref{fig:polysusc}). A CW fit in the same temperature range (15 - 300\,K) yielded an effective magnetic moment of 0.31\,$\mu_B$/Ir atom (C = 0.01228\,cm$^3$K/mol), which is smaller than the one of the single crystals (0.44\,$\mu_B$/Ir atom)\cite{Dey2016}. Furthermore, no evidence for  Y$_2$O$_3$ or other impurities was found in its XRD characterization (not shown).

\begin{figure}[b]
\centering
\includegraphics[width=\linewidth]{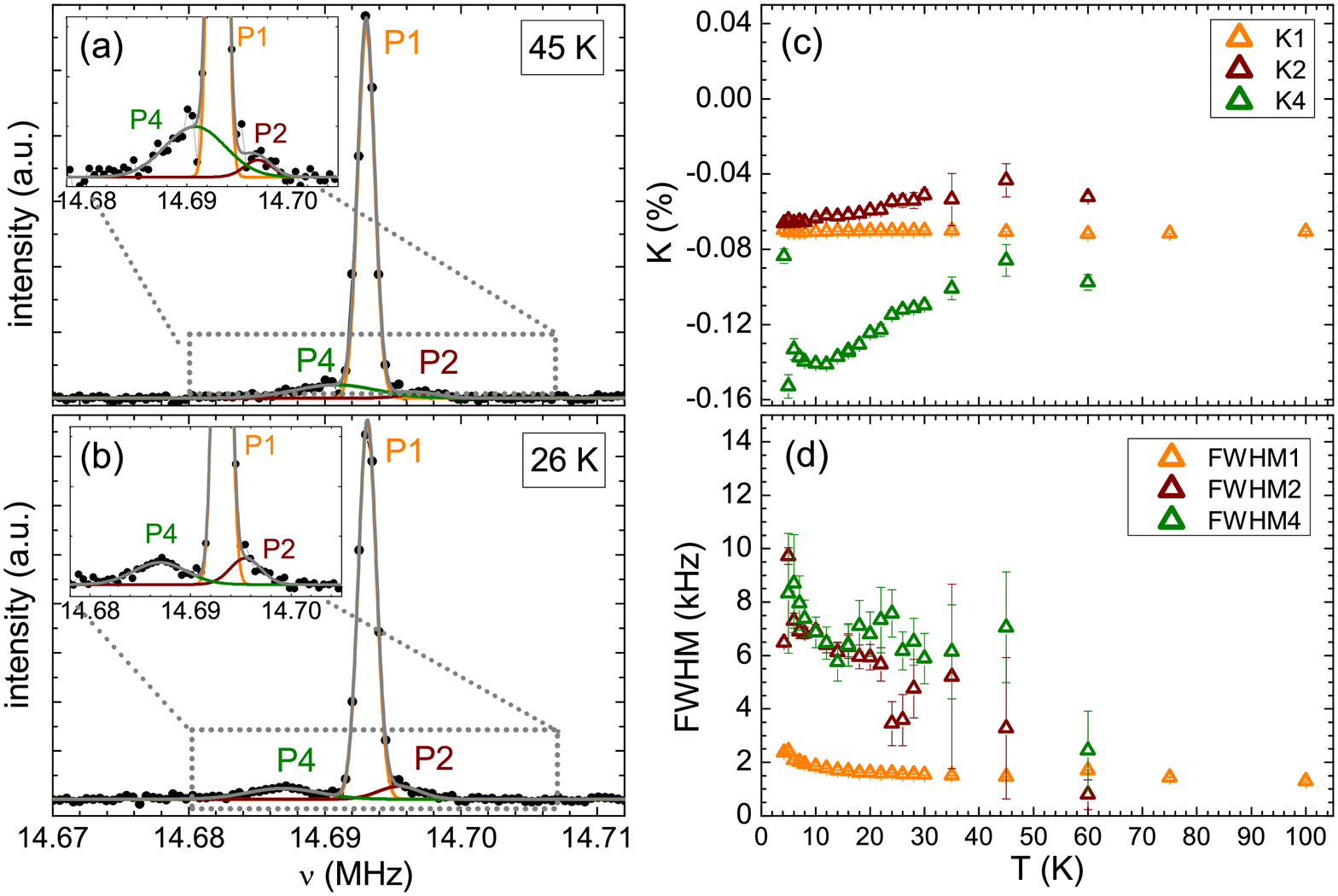}
\caption{\label{fig:NMRspectrapoly} (Color online) Left column (a), (b): Selected \y\ NMR spectra of the polycrystalline sample measured in a field of 7.0493\,T. Lines are Gaussian fits to the data, resulting in an anisotropic main line [peaks P1 (orange) and P2 (dark red)], as well as an additional peak at lower frequencies for the polycrystal (P4, green). 
Note that P3 (P4) is absent in the spectra of the polycrystal (single crystal), respectively. The insets in the left column show a zoom into the data to better reflect P2 and P4. Right column: Temperatur-dependent NMR Knight shifts (c) and FWHM (d) for the three NMR resonance lines (P1 - orange, P2 - dark red, and P4 - dark green) of the \bayiro\ polycrystal, measured in a field of 7.0493\,T.  Note that due to the reduced signal-to-noise ratio at high temperatures, peaks P2 and P4 could not be resolved at 75\, and 100\,K.} 
\end{figure}

\begin{figure}[t]
\centering
\includegraphics[width=0.8\linewidth]{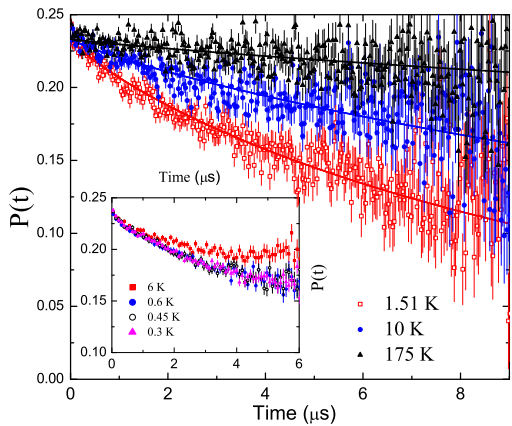}
\caption{\label{fig:ZF} (Color online) ZF $\mu$SR asymmetries of \bayiro\ measured at 175\,K (dark blue triangles), 10\,K (blue dots) and 1.51\,K (open red dots). Lines are fits to the data according to Eq.~\ref{eq:ZF}. The inset shows the evolution of the ZF $\mu$SR time spectra between 6\,K (closed red dots) and 300\,mK (pink triangles).} 
\end{figure}

This is reflected in the NMR spectra of the polycrystal [see Figs.~\ref{fig:NMRspectrapoly}(a) and (b)], where peak P3, which we assigned to nonmagnetic Y$_2$O$_3$ in the NMR spectra of the single crystals, is absent.
Furthermore, the spectral weight of P2, which presumably arises from chemical disorder and/or off-stoichiometry, is much smaller. An analysis of the spectral weight distribution between P1 and P2 yields a Y-Ir mismatch of about 1 - 1.5\,\% (see discussion of the spectral weights of the single crystal NMR). The main resonance line of the nonmagnetic \bayiro\ phase is found at the same frequency as for the single crystals and its Knight shift is the same as for the single crystal and is also temperature independent [see Fig.~\ref{fig:NMRspectrapoly}(c)]. A fit to $K_1(\chi)$ results in the same hyperfine coupling constant $A_{hf}$ and a very similar static shift $K_{stat}$ as for the single crystals [see Fig.~\ref{fig:hyperfine}]. Also its FWHM is very small (1.3\,kHz at 100\,K) and increases even less (up to 2.4\,kHz at 4.2\,K for P1 and $\sim$9\,kHz for P2) [Fig.~\ref{fig:NMRspectrapoly}(d)]. As for the single crystals, this is consistent with a calculated line broadening of $\Delta \nu$ = 6.5\,kHz due to the reduced magnetic moment of 0.31$\mu_B$/Ir atom (see discussion of the linewidth for the single crystals). Additionally, a small peak emerges at the lower frequency side (P4). This peak has a fast spin-lattice relaxation of roughly 100 - 300\,ms and its Knight shift shows a rather strong temperature dependence in comparison to the other two resonance lines [see Fig.~\ref{fig:NMRspectrapoly}(c)], suggesting that it emerges from Y nuclei sitting very close to a paramagnetic impurity, which in this case might be a tiny extrinsic paramagnetic cluster. (Note that this extrinsic phase also contributes to the measured macroscopic susceptibility. This means (i) the extrinsic paramagnetic phase is indeed very tiny and, (ii) the actual magnetic moment stemming from diluted paramagnetic electrons within the \bayiro\ main phase might even be smaller than what has been deduced from $\chi(T)$.) The spin-lattice relaxation time of P1 and P2 (which are indistinguishable) is very long, reaching the limits of a reasonable measurement duration. The \slrr\ data shown in the inset of Fig.~\ref{fig:NMRT1} are thus only rough estimates (note the big error bars on the logarithmic scale) and cannot be used for any further analysis. However, they suffice to point out the difference between the polycrystal  and the single crystals: The \slrr\ of the main peak of the polycrystal is at least one order of magnitude smaller than the one of the single crystals, stating that the less disordered polycrystal is even less magnetic than the single crystals. This confirms the nonmagnetic $J=0$ ground state of stoichiometric \bayiro\ and proofs that the weak magnetic signals arising in the magnetic susceptibility and in the NMR \slrr\ of the single crystals stem from a small amount of paramagnetic spin centers, presumably arising from a certain Y-Ir site disorder and/or off-stoichiometry within the \bayiro\ phase.

\subsection{Muon Spin Rotation}

Zero-field (ZF) $\mu$SR measurements have been carried out in the temperature range between 200\,K and 300\,mK. $\mu$SR time spectra at representative temperatures are shown in Fig.~\ref{fig:ZF}. Down to the lowest measured temperature of 300\,mK, the ZF $\mu$SR time spectra exhibit a weak exponential decay. No signal of static long-range 
magnetic order, or a fast muon relaxation associated with a strongly disordered magnetic state are observed, in accordance with different bulk characterization methods\cite{Dey2016} and NMR. 
The spectra have been fitted with a stretched exponential function:
\begin{equation}
A(t) = A_0 \exp(-\lambda t)^\beta      \label{eq:ZF}
\end{equation}
with the spin-lattice relaxation rate $\lambda$ and a stretched exponent $\beta$. The value of $\beta = 0.85$ close to one suggests that the system is rather homogeneous since for a strongly disordered system such as a spin glass or some other spin-frozen state the expected $\beta$ value should be much lower than 1 (i.e. close to 1/3 for a spin glass)\cite{Campbell1994}.

The spin-lattice relaxation rate $\lambda$ [see Fig.~\ref{fig:ZFT1}] remains temperature independent down to $\sim$ 60\,K, in conjunction with the magnetic susceptibility\cite{Dey2016} and the NMR \slrr. The non-zero value of $\lambda$ above 60\,K is associated with randomly distributed nuclear moments coupled with the muon spin. Below $\sim$ 60\,K, $\lambda$ starts to increase with decreasing temperature. This increase indicates that 
slow magnetic fluctuations develop, consistent with the increase of the \y\ NMR \slrr\ described above.
It should be noted that the absolute values of $\lambda$ in the temperature range from 60\,K down to 300\,mK are still very small, indicating that the overall density and size of the fluctuating localized moments are small. 
In contrast to the NMR \slrr\ the muon spin relaxation rate $\lambda$ continously increases down to 600\,mK. This is due to the different time windows of the electronic fluctuations, to which NMR and $\mu$SR are sensitive to. Below 600\,mK down to 300\,mK no further change of the muon spin relaxation is observed.

To clarify whether the low temperature muon spin relaxation is static or dynamic in nature, two decoupling experiments have been performed at 3.7\,K and 300\,mK in different longitudinal fields (LF $\mu$SR). 
At 3.7\,K one can completely suppress the muon spin relaxation with a 200\,G longitudinal field (not shown).
Whereas, at 300\,mK  [see inset of Fig.~\ref{fig:ZFT1}] even a longitudinal field of 100\,G is sufficient to decouple the muon spin relaxation. 

This proves that the weak magnetic field distribution is quasistatic with a fluctuation rate below 1\,MHz at 3.7\,K and 300\,mK. 
The data support the presence of diluted localized paramagnetic moments which exhibit a slowing down of fluctuations due to the presence of an energy barrier. This interpretation is consistent with the results of NMR (see above), specific heat and susceptibility measurements.\cite{Dey2016, Corredor2017}

\section{Conclusion}

By using local probe techniques, evidence for weak magnetic fluctuations in single crystals of \bayiro\ could be revealed. Instead of pointing towards an exotic quasi-static ground state of \bayiro, we could show that these fluctuations stem from tiny amounts of diluted paramagnetic impurity centers on the background of the nonmagnetic $J = 0$ ground state.
Due to the local character of NMR measurements, an impurity phase of nonmagnetic Y$_2$O$_3$ could be resolved in single crystals of \bayiro, in nice agreement with former XRD analysis. The local susceptibility of the intrinsic \bayiro\ phase as measured by the NMR Knight shift was found to be temperature independent, in contrast to bulk susceptibility measurements, which are influenced by tiny amounts of Y-Ir disorder and/or off-stoichiometry, leading to a certain amount of paramagnetic impurities. The existence of these paramagnetic spin centers is expressed in an additional peak (P2) of the NMR spectrum, an increase in its FWHM and a pronounced peak in \slrr\ at low temperature, indicating a progressive slowing down of weak magnetic fluctuations of these spin centers.  
These results are in perfect agreement with $\mu$SR measurements on the same samples, where a slight increase of the ZF $\mu$SR spin-lattice relaxation rate $\lambda$ and corresponding decoupling experiments reveal weak magnetic fluctuations. 
Comparing the activation energy of the magnetic fluctuations $E_a = (40 \pm 4)$\,K resulting from a BPP analysis of the NMR spin-lattice relaxation rate, and the onset temperature of the increase of the $\mu$SR spin-lattice relaxation rate with the theoretically and experimentally determined excitation gap between the $J = 0$ and $J = 1$ states of about 370\,meV in this compound, it is evident that the theoretically suggested excitonic type of magnetism cannot account for the observed magnetic response. Instead, we propose that these weak magnetic fluctuations stem from tiny amounts of diluted paramagnetic impurities on the background of a nonmagnetic $J = 0$ ground state. Measurements on a 
less disordered polycrystal of \bayiro\  further corroborated this scenario, since no evidence for magnetic fluctuations was found in this sample.\\

{\it Note added in proof} After submission of our manuscript, another work appeared which also discusses the effect of site disorder on the magnetism of \bayiro.\cite{Chen2017}

\section*{Acknowledgement}

We thank V. Kataev, L.~T.~Corredor-Bohorquez, G.~Prando, M.~Sturza and D.~Efremov for discussion and K.~Leger, S.~M\"uller-Litvyani, and J.~Werner for technical support. This work has been supported by the Deutsche Forschungsgemeinschaft (DFG) through the Sonderforschungsbereich SFB 1143, grant No. WO1532/3-2, the Emmy-Noether program (Grant No. WU595/3-1), and materials world network (WU595/5.1).


\begin{thebibliography}{48}
\expandafter\ifx\csname natexlab\endcsname\relax\def\natexlab#1{#1}\fi
\expandafter\ifx\csname bibnamefont\endcsname\relax
  \def\bibnamefont#1{#1}\fi
\expandafter\ifx\csname bibfnamefont\endcsname\relax
  \def\bibfnamefont#1{#1}\fi
\expandafter\ifx\csname citenamefont\endcsname\relax
  \def\citenamefont#1{#1}\fi
\expandafter\ifx\csname url\endcsname\relax
  \def\url#1{\texttt{#1}}\fi
\expandafter\ifx\csname urlprefix\endcsname\relax\def\urlprefix{URL }\fi
\providecommand{\bibinfo}[2]{#2}
\providecommand{\eprint}[2][]{\url{#2}}

\bibitem[{\citenamefont{Jackeli and Khaliullin}(2009)}]{Jackeli2009}
\bibinfo{author}{\bibfnamefont{G.}~\bibnamefont{Jackeli}} \bibnamefont{and}
  \bibinfo{author}{\bibfnamefont{G.}~\bibnamefont{Khaliullin}},
  \bibinfo{journal}{Phys. Rev. Lett.} \textbf{\bibinfo{volume}{102}},
  \bibinfo{pages}{017205} (\bibinfo{year}{2009}).

\bibitem[{\citenamefont{Chaloupka et~al.}(2010)\citenamefont{Chaloupka,
  Jackeli, and Khaliullin}}]{Chaloupka2010}
\bibinfo{author}{\bibfnamefont{J.~c.~v.} \bibnamefont{Chaloupka}},
  \bibinfo{author}{\bibfnamefont{G.}~\bibnamefont{Jackeli}}, \bibnamefont{and}
  \bibinfo{author}{\bibfnamefont{G.}~\bibnamefont{Khaliullin}},
  \bibinfo{journal}{Phys. Rev. Lett.} \textbf{\bibinfo{volume}{105}},
  \bibinfo{pages}{027204} (\bibinfo{year}{2010}).

\bibitem[{\citenamefont{Kim et~al.}(2008)\citenamefont{Kim, Jin, Moon, Kim,
  Park, Leem, Yu, Noh, Kim, Oh et~al.}}]{Kim2008PRL}
\bibinfo{author}{\bibfnamefont{B.~J.} \bibnamefont{Kim}},
  \bibinfo{author}{\bibfnamefont{H.}~\bibnamefont{Jin}},
  \bibinfo{author}{\bibfnamefont{S.~J.} \bibnamefont{Moon}},
  \bibinfo{author}{\bibfnamefont{J.-Y.} \bibnamefont{Kim}},
  \bibinfo{author}{\bibfnamefont{B.-G.} \bibnamefont{Park}},
  \bibinfo{author}{\bibfnamefont{C.~S.} \bibnamefont{Leem}},
  \bibinfo{author}{\bibfnamefont{J.}~\bibnamefont{Yu}},
  \bibinfo{author}{\bibfnamefont{T.~W.} \bibnamefont{Noh}},
  \bibinfo{author}{\bibfnamefont{C.}~\bibnamefont{Kim}},
  \bibinfo{author}{\bibfnamefont{S.-J.} \bibnamefont{Oh}},
  \bibnamefont{et~al.}, \bibinfo{journal}{Phys. Rev. Lett.}
  \textbf{\bibinfo{volume}{101}}, \bibinfo{pages}{076402}
  (\bibinfo{year}{2008}).

\bibitem[{\citenamefont{Okabe et~al.}(2011)\citenamefont{Okabe, Isobe,
  Takayama-Muromachi, Koda, Takeshita, Hiraishi, Miyazaki, Kadono, Miyake, and
  Akimitsu}}]{Okabe2011}
\bibinfo{author}{\bibfnamefont{H.}~\bibnamefont{Okabe}},
  \bibinfo{author}{\bibfnamefont{M.}~\bibnamefont{Isobe}},
  \bibinfo{author}{\bibfnamefont{E.}~\bibnamefont{Takayama-Muromachi}},
  \bibinfo{author}{\bibfnamefont{A.}~\bibnamefont{Koda}},
  \bibinfo{author}{\bibfnamefont{S.}~\bibnamefont{Takeshita}},
  \bibinfo{author}{\bibfnamefont{M.}~\bibnamefont{Hiraishi}},
  \bibinfo{author}{\bibfnamefont{M.}~\bibnamefont{Miyazaki}},
  \bibinfo{author}{\bibfnamefont{R.}~\bibnamefont{Kadono}},
  \bibinfo{author}{\bibfnamefont{Y.}~\bibnamefont{Miyake}}, \bibnamefont{and}
  \bibinfo{author}{\bibfnamefont{J.}~\bibnamefont{Akimitsu}},
  \bibinfo{journal}{Phys. Rev. B} \textbf{\bibinfo{volume}{83}},
  \bibinfo{pages}{155118} (\bibinfo{year}{2011}).

\bibitem[{\citenamefont{Boseggia et~al.}(2013)\citenamefont{Boseggia,
  Springell, Walker, R\o{}nnow, R\"uegg, Okabe, Isobe, Perry, Collins, and
  McMorrow}}]{Boseggia2013}
\bibinfo{author}{\bibfnamefont{S.}~\bibnamefont{Boseggia}},
  \bibinfo{author}{\bibfnamefont{R.}~\bibnamefont{Springell}},
  \bibinfo{author}{\bibfnamefont{H.~C.} \bibnamefont{Walker}},
  \bibinfo{author}{\bibfnamefont{H.~M.} \bibnamefont{R\o{}nnow}},
  \bibinfo{author}{\bibfnamefont{C.}~\bibnamefont{R\"uegg}},
  \bibinfo{author}{\bibfnamefont{H.}~\bibnamefont{Okabe}},
  \bibinfo{author}{\bibfnamefont{M.}~\bibnamefont{Isobe}},
  \bibinfo{author}{\bibfnamefont{R.~S.} \bibnamefont{Perry}},
  \bibinfo{author}{\bibfnamefont{S.~P.} \bibnamefont{Collins}},
  \bibnamefont{and} \bibinfo{author}{\bibfnamefont{D.~F.}
  \bibnamefont{McMorrow}}, \bibinfo{journal}{Phys. Rev. Lett.}
  \textbf{\bibinfo{volume}{110}}, \bibinfo{pages}{117207}
  (\bibinfo{year}{2013}).

\bibitem[{\citenamefont{Shitade et~al.}(2009)\citenamefont{Shitade, Katsura,
  Kune\ifmmode~\check{s}\else \v{s}\fi{}, Qi, Zhang, and
  Nagaosa}}]{Shitade2009}
\bibinfo{author}{\bibfnamefont{A.}~\bibnamefont{Shitade}},
  \bibinfo{author}{\bibfnamefont{H.}~\bibnamefont{Katsura}},
  \bibinfo{author}{\bibfnamefont{J.}~\bibnamefont{Kune\ifmmode~\check{s}\else
  \v{s}\fi{}}}, \bibinfo{author}{\bibfnamefont{X.-L.} \bibnamefont{Qi}},
  \bibinfo{author}{\bibfnamefont{S.-C.} \bibnamefont{Zhang}}, \bibnamefont{and}
  \bibinfo{author}{\bibfnamefont{N.}~\bibnamefont{Nagaosa}},
  \bibinfo{journal}{Phys. Rev. Lett.} \textbf{\bibinfo{volume}{102}},
  \bibinfo{pages}{256403} (\bibinfo{year}{2009}).

\bibitem[{\citenamefont{Singh and Gegenwart}(2010)}]{Singh2010}
\bibinfo{author}{\bibfnamefont{Y.}~\bibnamefont{Singh}} \bibnamefont{and}
  \bibinfo{author}{\bibfnamefont{P.}~\bibnamefont{Gegenwart}},
  \bibinfo{journal}{Phys. Rev. B} \textbf{\bibinfo{volume}{82}},
  \bibinfo{pages}{064412} (\bibinfo{year}{2010}).

\bibitem[{\citenamefont{Singh et~al.}(2012)\citenamefont{Singh, Manni, Reuther,
  Berlijn, Thomale, Ku, Trebst, and Gegenwart}}]{Singh2012}
\bibinfo{author}{\bibfnamefont{Y.}~\bibnamefont{Singh}},
  \bibinfo{author}{\bibfnamefont{S.}~\bibnamefont{Manni}},
  \bibinfo{author}{\bibfnamefont{J.}~\bibnamefont{Reuther}},
  \bibinfo{author}{\bibfnamefont{T.}~\bibnamefont{Berlijn}},
  \bibinfo{author}{\bibfnamefont{R.}~\bibnamefont{Thomale}},
  \bibinfo{author}{\bibfnamefont{W.}~\bibnamefont{Ku}},
  \bibinfo{author}{\bibfnamefont{S.}~\bibnamefont{Trebst}}, \bibnamefont{and}
  \bibinfo{author}{\bibfnamefont{P.}~\bibnamefont{Gegenwart}},
  \bibinfo{journal}{Phys. Rev. Lett.} \textbf{\bibinfo{volume}{108}},
  \bibinfo{pages}{127203} (\bibinfo{year}{2012}).

\bibitem[{\citenamefont{Choi et~al.}(2012)\citenamefont{Choi, Coldea,
  Kolmogorov, Lancaster, Mazin, Blundell, Radaelli, Singh, Gegenwart, Choi
  et~al.}}]{Choi2012}
\bibinfo{author}{\bibfnamefont{S.~K.} \bibnamefont{Choi}},
  \bibinfo{author}{\bibfnamefont{R.}~\bibnamefont{Coldea}},
  \bibinfo{author}{\bibfnamefont{A.~N.} \bibnamefont{Kolmogorov}},
  \bibinfo{author}{\bibfnamefont{T.}~\bibnamefont{Lancaster}},
  \bibinfo{author}{\bibfnamefont{I.~I.} \bibnamefont{Mazin}},
  \bibinfo{author}{\bibfnamefont{S.~J.} \bibnamefont{Blundell}},
  \bibinfo{author}{\bibfnamefont{P.~G.} \bibnamefont{Radaelli}},
  \bibinfo{author}{\bibfnamefont{Y.}~\bibnamefont{Singh}},
  \bibinfo{author}{\bibfnamefont{P.}~\bibnamefont{Gegenwart}},
  \bibinfo{author}{\bibfnamefont{K.~R.} \bibnamefont{Choi}},
  \bibnamefont{et~al.}, \bibinfo{journal}{Phys. Rev. Lett.}
  \textbf{\bibinfo{volume}{108}}, \bibinfo{pages}{127204}
  (\bibinfo{year}{2012}).

\bibitem[{\citenamefont{Gretarsson et~al.}(2013)\citenamefont{Gretarsson,
  Clancy, Liu, Hill, Bozin, Singh, Manni, Gegenwart, Kim, Said
  et~al.}}]{Gretarsson2013}
\bibinfo{author}{\bibfnamefont{H.}~\bibnamefont{Gretarsson}},
  \bibinfo{author}{\bibfnamefont{J.~P.} \bibnamefont{Clancy}},
  \bibinfo{author}{\bibfnamefont{X.}~\bibnamefont{Liu}},
  \bibinfo{author}{\bibfnamefont{J.~P.} \bibnamefont{Hill}},
  \bibinfo{author}{\bibfnamefont{E.}~\bibnamefont{Bozin}},
  \bibinfo{author}{\bibfnamefont{Y.}~\bibnamefont{Singh}},
  \bibinfo{author}{\bibfnamefont{S.}~\bibnamefont{Manni}},
  \bibinfo{author}{\bibfnamefont{P.}~\bibnamefont{Gegenwart}},
  \bibinfo{author}{\bibfnamefont{J.}~\bibnamefont{Kim}},
  \bibinfo{author}{\bibfnamefont{A.~H.} \bibnamefont{Said}},
  \bibnamefont{et~al.}, \bibinfo{journal}{Phys. Rev. Lett.}
  \textbf{\bibinfo{volume}{110}}, \bibinfo{pages}{076402}
  (\bibinfo{year}{2013}).

\bibitem[{\citenamefont{Cao et~al.}(2000)\citenamefont{Cao, Crow, Guertin,
  Henning, Homes, Strongin, Basov, and Lochner}}]{Cao2000}
\bibinfo{author}{\bibfnamefont{G.}~\bibnamefont{Cao}},
  \bibinfo{author}{\bibfnamefont{J.}~\bibnamefont{Crow}},
  \bibinfo{author}{\bibfnamefont{R.}~\bibnamefont{Guertin}},
  \bibinfo{author}{\bibfnamefont{P.}~\bibnamefont{Henning}},
  \bibinfo{author}{\bibfnamefont{C.}~\bibnamefont{Homes}},
  \bibinfo{author}{\bibfnamefont{M.}~\bibnamefont{Strongin}},
  \bibinfo{author}{\bibfnamefont{D.}~\bibnamefont{Basov}}, \bibnamefont{and}
  \bibinfo{author}{\bibfnamefont{E.}~\bibnamefont{Lochner}},
  \bibinfo{journal}{Solid State Communications} \textbf{\bibinfo{volume}{113}},
  \bibinfo{pages}{657 } (\bibinfo{year}{2000}).

\bibitem[{\citenamefont{Brooks et~al.}(2005)\citenamefont{Brooks, Blundell,
  Lancaster, Hayes, Pratt, Frampton, and Battle}}]{Brooks2005}
\bibinfo{author}{\bibfnamefont{M.~L.} \bibnamefont{Brooks}},
  \bibinfo{author}{\bibfnamefont{S.~J.} \bibnamefont{Blundell}},
  \bibinfo{author}{\bibfnamefont{T.}~\bibnamefont{Lancaster}},
  \bibinfo{author}{\bibfnamefont{W.}~\bibnamefont{Hayes}},
  \bibinfo{author}{\bibfnamefont{F.~L.} \bibnamefont{Pratt}},
  \bibinfo{author}{\bibfnamefont{P.~P.~C.} \bibnamefont{Frampton}},
  \bibnamefont{and} \bibinfo{author}{\bibfnamefont{P.~D.}
  \bibnamefont{Battle}}, \bibinfo{journal}{Phys. Rev. B}
  \textbf{\bibinfo{volume}{71}}, \bibinfo{pages}{220411}
  (\bibinfo{year}{2005}).

\bibitem[{\citenamefont{Khaliullin}(2013)}]{Khaliullin2013}
\bibinfo{author}{\bibfnamefont{G.}~\bibnamefont{Khaliullin}},
  \bibinfo{journal}{Phys. Rev. Lett.} \textbf{\bibinfo{volume}{111}},
  \bibinfo{pages}{197201} (\bibinfo{year}{2013}).

\bibitem[{\citenamefont{Meetei et~al.}(2015)\citenamefont{Meetei, Cole,
  Randeria, and Trivedi}}]{Meetei2015}
\bibinfo{author}{\bibfnamefont{O.~N.} \bibnamefont{Meetei}},
  \bibinfo{author}{\bibfnamefont{W.~S.} \bibnamefont{Cole}},
  \bibinfo{author}{\bibfnamefont{M.}~\bibnamefont{Randeria}}, \bibnamefont{and}
  \bibinfo{author}{\bibfnamefont{N.}~\bibnamefont{Trivedi}},
  \bibinfo{journal}{Phys. Rev. B} \textbf{\bibinfo{volume}{91}},
  \bibinfo{pages}{054412} (\bibinfo{year}{2015}).

\bibitem[{\citenamefont{Cao et~al.}(2014)\citenamefont{Cao, Qi, Li, Terzic,
  Yuan, DeLong, Murthy, and Kaul}}]{Cao2014}
\bibinfo{author}{\bibfnamefont{G.}~\bibnamefont{Cao}},
  \bibinfo{author}{\bibfnamefont{T.}~\bibnamefont{Qi}},
  \bibinfo{author}{\bibfnamefont{L.}~\bibnamefont{Li}},
  \bibinfo{author}{\bibfnamefont{J.}~\bibnamefont{Terzic}},
  \bibinfo{author}{\bibfnamefont{S.}~\bibnamefont{Yuan}},
  \bibinfo{author}{\bibfnamefont{L.}~\bibnamefont{DeLong}},
  \bibinfo{author}{\bibfnamefont{G.}~\bibnamefont{Murthy}}, \bibnamefont{and}
  \bibinfo{author}{\bibfnamefont{R.}~\bibnamefont{Kaul}},
  \bibinfo{journal}{Phys. Rev. Lett.} \textbf{\bibinfo{volume}{112}},
  \bibinfo{pages}{056402} (\bibinfo{year}{2014}).

\bibitem[{\citenamefont{Terizc et~al.}(2016)\citenamefont{Terizc, Zheng, Ye,
  Zhao, Schlottmann, Long, and Cao}}]{Terizc2016}
\bibinfo{author}{\bibfnamefont{J.}~\bibnamefont{Terizc}},
  \bibinfo{author}{\bibfnamefont{H.}~\bibnamefont{Zheng}},
  \bibinfo{author}{\bibfnamefont{F.}~\bibnamefont{Ye}},
  \bibinfo{author}{\bibfnamefont{H.~D.} \bibnamefont{Zhao}},
  \bibinfo{author}{\bibfnamefont{P.}~\bibnamefont{Schlottmann}},
  \bibinfo{author}{\bibfnamefont{L.~D.} \bibnamefont{Long}}, \bibnamefont{and}
  \bibinfo{author}{\bibfnamefont{G.}~\bibnamefont{Cao}} (\bibinfo{year}{2016}),
  \eprint{arXiv:1608.07624}.

\bibitem[{\citenamefont{Cao et~al.}(2004)\citenamefont{Cao, Lin, Chikara,
  Durairaj, and Elhami}}]{Cao2004}
\bibinfo{author}{\bibfnamefont{G.}~\bibnamefont{Cao}},
  \bibinfo{author}{\bibfnamefont{X.~N.} \bibnamefont{Lin}},
  \bibinfo{author}{\bibfnamefont{S.}~\bibnamefont{Chikara}},
  \bibinfo{author}{\bibfnamefont{V.}~\bibnamefont{Durairaj}}, \bibnamefont{and}
  \bibinfo{author}{\bibfnamefont{E.}~\bibnamefont{Elhami}},
  \bibinfo{journal}{Phys. Rev. B} \textbf{\bibinfo{volume}{69}},
  \bibinfo{pages}{174418} (\bibinfo{year}{2004}).

\bibitem[{\citenamefont{Yuan et~al.}(2016)\citenamefont{Yuan, Butrouna, Terzic,
  Zheng, Aswartham, DeLong, Ye, Schlottmann, and Cao}}]{Yuan2016}
\bibinfo{author}{\bibfnamefont{S.~J.} \bibnamefont{Yuan}},
  \bibinfo{author}{\bibfnamefont{K.}~\bibnamefont{Butrouna}},
  \bibinfo{author}{\bibfnamefont{J.}~\bibnamefont{Terzic}},
  \bibinfo{author}{\bibfnamefont{H.}~\bibnamefont{Zheng}},
  \bibinfo{author}{\bibfnamefont{S.}~\bibnamefont{Aswartham}},
  \bibinfo{author}{\bibfnamefont{L.~E.} \bibnamefont{DeLong}},
  \bibinfo{author}{\bibfnamefont{F.}~\bibnamefont{Ye}},
  \bibinfo{author}{\bibfnamefont{P.}~\bibnamefont{Schlottmann}},
  \bibnamefont{and} \bibinfo{author}{\bibfnamefont{G.}~\bibnamefont{Cao}},
  \bibinfo{journal}{Phys. Rev. B} \textbf{\bibinfo{volume}{93}},
  \bibinfo{pages}{165136} (\bibinfo{year}{2016}).

\bibitem[{\citenamefont{Ranjbar et~al.}(2015)\citenamefont{Ranjbar, Reynolds,
  Kayser, Kennedy, Hester, and Kimpton}}]{Ranjbar2015}
\bibinfo{author}{\bibfnamefont{B.}~\bibnamefont{Ranjbar}},
  \bibinfo{author}{\bibfnamefont{E.}~\bibnamefont{Reynolds}},
  \bibinfo{author}{\bibfnamefont{P.}~\bibnamefont{Kayser}},
  \bibinfo{author}{\bibfnamefont{B.~J.} \bibnamefont{Kennedy}},
  \bibinfo{author}{\bibfnamefont{J.~R.} \bibnamefont{Hester}},
  \bibnamefont{and} \bibinfo{author}{\bibfnamefont{J.~A.}
  \bibnamefont{Kimpton}}, \bibinfo{journal}{Inorganic Chemistry}
  \textbf{\bibinfo{volume}{54}}, \bibinfo{pages}{10468} (\bibinfo{year}{2015}).

\bibitem[{\citenamefont{Phelan et~al.}(2016)\citenamefont{Phelan, Seibel, Jr.,
  Xie, and Cava}}]{Phelan2016}
\bibinfo{author}{\bibfnamefont{B.~F.} \bibnamefont{Phelan}},
  \bibinfo{author}{\bibfnamefont{E.~M.} \bibnamefont{Seibel}},
  \bibinfo{author}{\bibfnamefont{D.~B.} \bibnamefont{Jr.}},
  \bibinfo{author}{\bibfnamefont{W.}~\bibnamefont{Xie}}, \bibnamefont{and}
  \bibinfo{author}{\bibfnamefont{R.}~\bibnamefont{Cava}},
  \bibinfo{journal}{Solid State Communications} \textbf{\bibinfo{volume}{236}},
  \bibinfo{pages}{37 } (\bibinfo{year}{2016}).

\bibitem[{\citenamefont{Wakeshima et~al.}(1999)\citenamefont{Wakeshima, Harada,
  and Hinatsu}}]{Wakeshima1999}
\bibinfo{author}{\bibfnamefont{M.}~\bibnamefont{Wakeshima}},
  \bibinfo{author}{\bibfnamefont{D.}~\bibnamefont{Harada}}, \bibnamefont{and}
  \bibinfo{author}{\bibfnamefont{Y.}~\bibnamefont{Hinatsu}},
  \bibinfo{journal}{Journal of Alloys and Compounds}
  \textbf{\bibinfo{volume}{287}}, \bibinfo{pages}{130 } (\bibinfo{year}{1999}).

\bibitem[{\citenamefont{Corredor et~al.}(2017)\citenamefont{Corredor,
  Aslan-Cansever, Sturza, Manna, Maljuk, Gass, Dey, Wolter, Kataeva, Zimmermann
  et~al.}}]{Corredor2017}
\bibinfo{author}{\bibfnamefont{L.~T.} \bibnamefont{Corredor}},
  \bibinfo{author}{\bibfnamefont{G.}~\bibnamefont{Aslan-Cansever}},
  \bibinfo{author}{\bibfnamefont{M.}~\bibnamefont{Sturza}},
  \bibinfo{author}{\bibfnamefont{K.}~\bibnamefont{Manna}},
  \bibinfo{author}{\bibfnamefont{A.}~\bibnamefont{Maljuk}},
  \bibinfo{author}{\bibfnamefont{S.}~\bibnamefont{Gass}},
  \bibinfo{author}{\bibfnamefont{T.}~\bibnamefont{Dey}},
  \bibinfo{author}{\bibfnamefont{A.~U.~B.} \bibnamefont{Wolter}},
  \bibinfo{author}{\bibfnamefont{O.}~\bibnamefont{Kataeva}},
  \bibinfo{author}{\bibfnamefont{A.}~\bibnamefont{Zimmermann}},
  \bibnamefont{et~al.}, \bibinfo{journal}{Phys. Rev. B}
  \textbf{\bibinfo{volume}{95}}, \bibinfo{pages}{064418}
  (\bibinfo{year}{2017}).

\bibitem[{\citenamefont{Dey et~al.}(2016)\citenamefont{Dey, Maljuk, Efremov,
  Kataeva, Gass, Blum, Steckel, Gruner, Ritschel, Wolter et~al.}}]{Dey2016}
\bibinfo{author}{\bibfnamefont{T.}~\bibnamefont{Dey}},
  \bibinfo{author}{\bibfnamefont{A.}~\bibnamefont{Maljuk}},
  \bibinfo{author}{\bibfnamefont{D.~V.} \bibnamefont{Efremov}},
  \bibinfo{author}{\bibfnamefont{O.}~\bibnamefont{Kataeva}},
  \bibinfo{author}{\bibfnamefont{S.}~\bibnamefont{Gass}},
  \bibinfo{author}{\bibfnamefont{C.~G.~F.} \bibnamefont{Blum}},
  \bibinfo{author}{\bibfnamefont{F.}~\bibnamefont{Steckel}},
  \bibinfo{author}{\bibfnamefont{D.}~\bibnamefont{Gruner}},
  \bibinfo{author}{\bibfnamefont{T.}~\bibnamefont{Ritschel}},
  \bibinfo{author}{\bibfnamefont{A.~U.~B.} \bibnamefont{Wolter}},
  \bibnamefont{et~al.}, \bibinfo{journal}{Phys. Rev. B}
  \textbf{\bibinfo{volume}{93}}, \bibinfo{pages}{014434}
  (\bibinfo{year}{2016}).

\bibitem[{\citenamefont{Johnston et~al.}(2005)\citenamefont{Johnston, Baek,
  Zong, Borsa, Schmalian, and Kondo}}]{Johnston2005}
\bibinfo{author}{\bibfnamefont{D.~C.} \bibnamefont{Johnston}},
  \bibinfo{author}{\bibfnamefont{S.-H.} \bibnamefont{Baek}},
  \bibinfo{author}{\bibfnamefont{X.}~\bibnamefont{Zong}},
  \bibinfo{author}{\bibfnamefont{F.}~\bibnamefont{Borsa}},
  \bibinfo{author}{\bibfnamefont{J.}~\bibnamefont{Schmalian}},
  \bibnamefont{and} \bibinfo{author}{\bibfnamefont{S.}~\bibnamefont{Kondo}},
  \bibinfo{journal}{Phys. Rev. Lett.} \textbf{\bibinfo{volume}{95}},
  \bibinfo{pages}{176408} (\bibinfo{year}{2005}).

\bibitem[{\citenamefont{Johnston}(2006)}]{Johnston2006}
\bibinfo{author}{\bibfnamefont{D.~C.} \bibnamefont{Johnston}},
  \bibinfo{journal}{Phys. Rev. B} \textbf{\bibinfo{volume}{74}},
  \bibinfo{pages}{184430} (\bibinfo{year}{2006}).

\bibitem[{\citenamefont{Shiroka et~al.}(2011)\citenamefont{Shiroka, Casola,
  Glazkov, Zheludev, Pr\ifmmode~\check{s}\else \v{s}\fi{}a, Ott, and
  Mesot}}]{Shiroka2011}
\bibinfo{author}{\bibfnamefont{T.}~\bibnamefont{Shiroka}},
  \bibinfo{author}{\bibfnamefont{F.}~\bibnamefont{Casola}},
  \bibinfo{author}{\bibfnamefont{V.}~\bibnamefont{Glazkov}},
  \bibinfo{author}{\bibfnamefont{A.}~\bibnamefont{Zheludev}},
  \bibinfo{author}{\bibfnamefont{K.}~\bibnamefont{Pr\ifmmode~\check{s}\else
  \v{s}\fi{}a}}, \bibinfo{author}{\bibfnamefont{H.-R.} \bibnamefont{Ott}},
  \bibnamefont{and} \bibinfo{author}{\bibfnamefont{J.}~\bibnamefont{Mesot}},
  \bibinfo{journal}{Phys. Rev. Lett.} \textbf{\bibinfo{volume}{106}},
  \bibinfo{pages}{137202} (\bibinfo{year}{2011}).

\bibitem[{\citenamefont{Mitrovi\ifmmode~\acute{c}\else \'{c}\fi{}
  et~al.}(2008)\citenamefont{Mitrovi\ifmmode~\acute{c}\else \'{c}\fi{}, Julien,
  de~Vaulx, Horvati\ifmmode~\acute{c}\else \'{c}\fi{}, Berthier, Suzuki, and
  Yamada}}]{Mitrovic2008}
\bibinfo{author}{\bibfnamefont{V.~F.}
  \bibnamefont{Mitrovi\ifmmode~\acute{c}\else \'{c}\fi{}}},
  \bibinfo{author}{\bibfnamefont{M.-H.} \bibnamefont{Julien}},
  \bibinfo{author}{\bibfnamefont{C.}~\bibnamefont{de~Vaulx}},
  \bibinfo{author}{\bibfnamefont{M.}~\bibnamefont{Horvati\ifmmode~\acute{c}\el%
se \'{c}\fi{}}}, \bibinfo{author}{\bibfnamefont{C.}~\bibnamefont{Berthier}},
  \bibinfo{author}{\bibfnamefont{T.}~\bibnamefont{Suzuki}}, \bibnamefont{and}
  \bibinfo{author}{\bibfnamefont{K.}~\bibnamefont{Yamada}},
  \bibinfo{journal}{Phys. Rev. B} \textbf{\bibinfo{volume}{78}},
  \bibinfo{pages}{014504} (\bibinfo{year}{2008}).

\bibitem[{\citenamefont{Dioguardi et~al.}(2015)\citenamefont{Dioguardi, Lawson,
  Bush, Crocker, Shirer, Nisson, Kissikov, Ran, Bud'ko, Canfield
  et~al.}}]{Dioguardi2015}
\bibinfo{author}{\bibfnamefont{A.~P.} \bibnamefont{Dioguardi}},
  \bibinfo{author}{\bibfnamefont{M.~M.} \bibnamefont{Lawson}},
  \bibinfo{author}{\bibfnamefont{B.~T.} \bibnamefont{Bush}},
  \bibinfo{author}{\bibfnamefont{J.}~\bibnamefont{Crocker}},
  \bibinfo{author}{\bibfnamefont{K.~R.} \bibnamefont{Shirer}},
  \bibinfo{author}{\bibfnamefont{D.~M.} \bibnamefont{Nisson}},
  \bibinfo{author}{\bibfnamefont{T.}~\bibnamefont{Kissikov}},
  \bibinfo{author}{\bibfnamefont{S.}~\bibnamefont{Ran}},
  \bibinfo{author}{\bibfnamefont{S.~L.} \bibnamefont{Bud'ko}},
  \bibinfo{author}{\bibfnamefont{P.~C.} \bibnamefont{Canfield}},
  \bibnamefont{et~al.}, \bibinfo{journal}{Phys. Rev. B}
  \textbf{\bibinfo{volume}{92}}, \bibinfo{pages}{165116}
  (\bibinfo{year}{2015}).

\bibitem[{\citenamefont{Schenck}(1985)}]{Schenck1985}
\bibinfo{author}{\bibfnamefont{A.}~\bibnamefont{Schenck}},
  \emph{\bibinfo{title}{Muon spin rotation spectroscopy}}
  (\bibinfo{year}{1985}).

\bibitem[{\citenamefont{Yaouanc and de~Réotier}(2011)}]{Yaouanc2011}
\bibinfo{author}{\bibfnamefont{A.}~\bibnamefont{Yaouanc}} \bibnamefont{and}
  \bibinfo{author}{\bibfnamefont{P.}~\bibnamefont{de~Réotier}},
  \emph{\bibinfo{title}{Muon Spin Rotation, Relaxation, and Resonance:
  Applications to Condensed Matter}}, International Series of Monographs on
  Physics (\bibinfo{publisher}{OUP Oxford}, \bibinfo{year}{2011}).

\bibitem[{\citenamefont{Harazono and Watanabe}(1997)}]{Harazono1997}
\bibinfo{author}{\bibfnamefont{T.}~\bibnamefont{Harazono}} \bibnamefont{and}
  \bibinfo{author}{\bibfnamefont{T.}~\bibnamefont{Watanabe}},
  \bibinfo{journal}{Bulletin of the Chemical Society of Japan}
  \textbf{\bibinfo{volume}{70}}, \bibinfo{pages}{2383} (\bibinfo{year}{1997}).

\bibitem[{\citenamefont{Johnston}(2010)}]{Johnston2010}
\bibinfo{author}{\bibfnamefont{D.~C.} \bibnamefont{Johnston}},
  \bibinfo{journal}{Advances in Physics} \textbf{\bibinfo{volume}{59}},
  \bibinfo{pages}{803} (\bibinfo{year}{2010}).

\bibitem[{\citenamefont{Demazeau et~al.}(1993)\citenamefont{Demazeau, Jung,
  Sanchez, Colineau, Blaise, and Fournes}}]{Demazeau1993}
\bibinfo{author}{\bibfnamefont{G.}~\bibnamefont{Demazeau}},
  \bibinfo{author}{\bibfnamefont{D.-Y.} \bibnamefont{Jung}},
  \bibinfo{author}{\bibfnamefont{J.-P.} \bibnamefont{Sanchez}},
  \bibinfo{author}{\bibfnamefont{E.}~\bibnamefont{Colineau}},
  \bibinfo{author}{\bibfnamefont{A.}~\bibnamefont{Blaise}}, \bibnamefont{and}
  \bibinfo{author}{\bibfnamefont{L.}~\bibnamefont{Fournes}},
  \bibinfo{journal}{Solid State Communications} \textbf{\bibinfo{volume}{85}},
  \bibinfo{pages}{479 } (\bibinfo{year}{1993}).

\bibitem[{\citenamefont{Aharen et~al.}(2009)\citenamefont{Aharen, Greedan,
  Ning, Imai, Michaelis, Kroeker, Zhou, Wiebe, and Cranswick}}]{Aharen2009}
\bibinfo{author}{\bibfnamefont{T.}~\bibnamefont{Aharen}},
  \bibinfo{author}{\bibfnamefont{J.~E.} \bibnamefont{Greedan}},
  \bibinfo{author}{\bibfnamefont{F.}~\bibnamefont{Ning}},
  \bibinfo{author}{\bibfnamefont{T.}~\bibnamefont{Imai}},
  \bibinfo{author}{\bibfnamefont{V.}~\bibnamefont{Michaelis}},
  \bibinfo{author}{\bibfnamefont{S.}~\bibnamefont{Kroeker}},
  \bibinfo{author}{\bibfnamefont{H.}~\bibnamefont{Zhou}},
  \bibinfo{author}{\bibfnamefont{C.~R.} \bibnamefont{Wiebe}}, \bibnamefont{and}
  \bibinfo{author}{\bibfnamefont{L.~M.~D.} \bibnamefont{Cranswick}},
  \bibinfo{journal}{Phys. Rev. B} \textbf{\bibinfo{volume}{80}},
  \bibinfo{pages}{134423} (\bibinfo{year}{2009}).

\bibitem[{\citenamefont{Aharen et~al.}(2010)\citenamefont{Aharen, Greedan,
  Bridges, Aczel, Rodriguez, MacDougall, Luke, Imai, Michaelis, Kroeker
  et~al.}}]{Aharen2010}
\bibinfo{author}{\bibfnamefont{T.}~\bibnamefont{Aharen}},
  \bibinfo{author}{\bibfnamefont{J.~E.} \bibnamefont{Greedan}},
  \bibinfo{author}{\bibfnamefont{C.~A.} \bibnamefont{Bridges}},
  \bibinfo{author}{\bibfnamefont{A.~A.} \bibnamefont{Aczel}},
  \bibinfo{author}{\bibfnamefont{J.}~\bibnamefont{Rodriguez}},
  \bibinfo{author}{\bibfnamefont{G.}~\bibnamefont{MacDougall}},
  \bibinfo{author}{\bibfnamefont{G.~M.} \bibnamefont{Luke}},
  \bibinfo{author}{\bibfnamefont{T.}~\bibnamefont{Imai}},
  \bibinfo{author}{\bibfnamefont{V.~K.} \bibnamefont{Michaelis}},
  \bibinfo{author}{\bibfnamefont{S.}~\bibnamefont{Kroeker}},
  \bibnamefont{et~al.}, \bibinfo{journal}{Phys. Rev. B}
  \textbf{\bibinfo{volume}{81}}, \bibinfo{pages}{224409}
  (\bibinfo{year}{2010}).

\bibitem[{rem()}]{remark}
\emph{\bibinfo{title}{One may wonder why the stretching exponent $\beta$ does
  not change in the temperature regime where the BPP peak of $T_1^{-1}$ occurs.
  However, $\beta \sim 0.5$ already points towards a rather broad distribution
  of relaxation rates, including the relaxation rates in the peak region.}}

\bibitem[{\citenamefont{Bloembergen et~al.}(1947)\citenamefont{Bloembergen,
  Purcell, and Pound}}]{BloembergenNature1947}
\bibinfo{author}{\bibfnamefont{N.}~\bibnamefont{Bloembergen}},
  \bibinfo{author}{\bibfnamefont{E.~M.} \bibnamefont{Purcell}},
  \bibnamefont{and} \bibinfo{author}{\bibfnamefont{R.~V.} \bibnamefont{Pound}},
  \bibinfo{journal}{Nature} \textbf{\bibinfo{volume}{160}},
  \bibinfo{pages}{475} (\bibinfo{year}{1947}).

\bibitem[{\citenamefont{Bloembergen et~al.}(1948)\citenamefont{Bloembergen,
  Purcell, and Pound}}]{Bloembergen1948}
\bibinfo{author}{\bibfnamefont{N.}~\bibnamefont{Bloembergen}},
  \bibinfo{author}{\bibfnamefont{E.~M.} \bibnamefont{Purcell}},
  \bibnamefont{and} \bibinfo{author}{\bibfnamefont{R.~V.} \bibnamefont{Pound}},
  \bibinfo{journal}{Phys. Rev.} \textbf{\bibinfo{volume}{73}},
  \bibinfo{pages}{679} (\bibinfo{year}{1948}).

\bibitem[{\citenamefont{Suh et~al.}(2000)\citenamefont{Suh, Hammel, H\"ucker,
  B\"uchner, Ammerahl, and Revcolevschi}}]{Suh2000}
\bibinfo{author}{\bibfnamefont{B.~J.} \bibnamefont{Suh}},
  \bibinfo{author}{\bibfnamefont{P.~C.} \bibnamefont{Hammel}},
  \bibinfo{author}{\bibfnamefont{M.}~\bibnamefont{H\"ucker}},
  \bibinfo{author}{\bibfnamefont{B.}~\bibnamefont{B\"uchner}},
  \bibinfo{author}{\bibfnamefont{U.}~\bibnamefont{Ammerahl}}, \bibnamefont{and}
  \bibinfo{author}{\bibfnamefont{A.}~\bibnamefont{Revcolevschi}},
  \bibinfo{journal}{Phys. Rev. B} \textbf{\bibinfo{volume}{61}},
  \bibinfo{pages}{R9265} (\bibinfo{year}{2000}).

\bibitem[{\citenamefont{Curro et~al.}(2000)\citenamefont{Curro, Hammel, Suh,
  H\"ucker, B\"uchner, Ammerahl, and Revcolevschi}}]{Curro2000}
\bibinfo{author}{\bibfnamefont{N.~J.} \bibnamefont{Curro}},
  \bibinfo{author}{\bibfnamefont{P.~C.} \bibnamefont{Hammel}},
  \bibinfo{author}{\bibfnamefont{B.~J.} \bibnamefont{Suh}},
  \bibinfo{author}{\bibfnamefont{M.}~\bibnamefont{H\"ucker}},
  \bibinfo{author}{\bibfnamefont{B.}~\bibnamefont{B\"uchner}},
  \bibinfo{author}{\bibfnamefont{U.}~\bibnamefont{Ammerahl}}, \bibnamefont{and}
  \bibinfo{author}{\bibfnamefont{A.}~\bibnamefont{Revcolevschi}},
  \bibinfo{journal}{Phys. Rev. Lett.} \textbf{\bibinfo{volume}{85}},
  \bibinfo{pages}{642} (\bibinfo{year}{2000}).

\bibitem[{\citenamefont{Curro et~al.}(2009)\citenamefont{Curro, Dioguardi,
  ApRoberts-Warren, Shockley, and Klavins}}]{Curro2009}
\bibinfo{author}{\bibfnamefont{N.~J.} \bibnamefont{Curro}},
  \bibinfo{author}{\bibfnamefont{A.~P.} \bibnamefont{Dioguardi}},
  \bibinfo{author}{\bibfnamefont{N.}~\bibnamefont{ApRoberts-Warren}},
  \bibinfo{author}{\bibfnamefont{A.~C.} \bibnamefont{Shockley}},
  \bibnamefont{and} \bibinfo{author}{\bibfnamefont{P.}~\bibnamefont{Klavins}},
  \bibinfo{journal}{New J. Phys.} \textbf{\bibinfo{volume}{11}},
  \bibinfo{pages}{075004} (\bibinfo{year}{2009}).

\bibitem[{\citenamefont{Hammerath et~al.}(2013)\citenamefont{Hammerath,
  Gr\"afe, K\"uhne, K\"uhne, Kuhns, Reyes, Lang, Wurmehl, B\"uchner, Carretta
  et~al.}}]{Hammerath2013}
\bibinfo{author}{\bibfnamefont{F.}~\bibnamefont{Hammerath}},
  \bibinfo{author}{\bibfnamefont{U.}~\bibnamefont{Gr\"afe}},
  \bibinfo{author}{\bibfnamefont{T.}~\bibnamefont{K\"uhne}},
  \bibinfo{author}{\bibfnamefont{H.}~\bibnamefont{K\"uhne}},
  \bibinfo{author}{\bibfnamefont{P.~L.} \bibnamefont{Kuhns}},
  \bibinfo{author}{\bibfnamefont{A.~P.} \bibnamefont{Reyes}},
  \bibinfo{author}{\bibfnamefont{G.}~\bibnamefont{Lang}},
  \bibinfo{author}{\bibfnamefont{S.}~\bibnamefont{Wurmehl}},
  \bibinfo{author}{\bibfnamefont{B.}~\bibnamefont{B\"uchner}},
  \bibinfo{author}{\bibfnamefont{P.}~\bibnamefont{Carretta}},
  \bibnamefont{et~al.}, \bibinfo{journal}{Phys. Rev. B}
  \textbf{\bibinfo{volume}{88}}, \bibinfo{pages}{104503}
  (\bibinfo{year}{2013}).

\bibitem[{\citenamefont{Hammerath et~al.}(2015)\citenamefont{Hammerath, Moroni,
  Bossoni, Sanna, Kappenberger, Wurmehl, Wolter, Afrassa, Kobayashi, Sato
  et~al.}}]{Hammerath2015}
\bibinfo{author}{\bibfnamefont{F.}~\bibnamefont{Hammerath}},
  \bibinfo{author}{\bibfnamefont{M.}~\bibnamefont{Moroni}},
  \bibinfo{author}{\bibfnamefont{L.}~\bibnamefont{Bossoni}},
  \bibinfo{author}{\bibfnamefont{S.}~\bibnamefont{Sanna}},
  \bibinfo{author}{\bibfnamefont{R.}~\bibnamefont{Kappenberger}},
  \bibinfo{author}{\bibfnamefont{S.}~\bibnamefont{Wurmehl}},
  \bibinfo{author}{\bibfnamefont{A.~U.~B.} \bibnamefont{Wolter}},
  \bibinfo{author}{\bibfnamefont{M.~A.} \bibnamefont{Afrassa}},
  \bibinfo{author}{\bibfnamefont{Y.}~\bibnamefont{Kobayashi}},
  \bibinfo{author}{\bibfnamefont{M.}~\bibnamefont{Sato}}, \bibnamefont{et~al.},
  \bibinfo{journal}{Phys. Rev. B} \textbf{\bibinfo{volume}{92}},
  \bibinfo{pages}{020505} (\bibinfo{year}{2015}).

\bibitem[{\citenamefont{Kusch et~al.}(2017)\citenamefont{Kusch, Katukuri,
  B\"uchner, Dey, Efremov, Hamann-Borrero, Kim, Krisch, Malyuk, Moretti
  et~al.}}]{Kusch2017}
\bibinfo{author}{\bibfnamefont{M.}~\bibnamefont{Kusch}},
  \bibinfo{author}{\bibfnamefont{V.~M.} \bibnamefont{Katukuri}},
  \bibinfo{author}{\bibfnamefont{B.}~\bibnamefont{B\"uchner}},
  \bibinfo{author}{\bibfnamefont{T.}~\bibnamefont{Dey}},
  \bibinfo{author}{\bibfnamefont{D.}~\bibnamefont{Efremov}},
  \bibinfo{author}{\bibfnamefont{J.~E.} \bibnamefont{Hamann-Borrero}},
  \bibinfo{author}{\bibfnamefont{B.~H.} \bibnamefont{Kim}},
  \bibinfo{author}{\bibfnamefont{M.}~\bibnamefont{Krisch}},
  \bibinfo{author}{\bibfnamefont{A.}~\bibnamefont{Malyuk}},
  \bibinfo{author}{\bibfnamefont{M.}~\bibnamefont{Moretti}},
  \bibnamefont{et~al.} (\bibinfo{year}{2017}).

\bibitem[{\citenamefont{Nag et~al.}(2017)\citenamefont{Nag, Bhowal, Sala,
  Efimenko, Bert, Biswas, Hillier, Itoh, Kaushik, Siruguri et~al.}}]{Nag2017}
\bibinfo{author}{\bibfnamefont{A.}~\bibnamefont{Nag}},
  \bibinfo{author}{\bibfnamefont{S.}~\bibnamefont{Bhowal}},
  \bibinfo{author}{\bibfnamefont{M.~M.} \bibnamefont{Sala}},
  \bibinfo{author}{\bibfnamefont{A.}~\bibnamefont{Efimenko}},
  \bibinfo{author}{\bibfnamefont{F.}~\bibnamefont{Bert}},
  \bibinfo{author}{\bibfnamefont{P.~K.} \bibnamefont{Biswas}},
  \bibinfo{author}{\bibfnamefont{A.~D.} \bibnamefont{Hillier}},
  \bibinfo{author}{\bibfnamefont{M.}~\bibnamefont{Itoh}},
  \bibinfo{author}{\bibfnamefont{S.~D.} \bibnamefont{Kaushik}},
  \bibinfo{author}{\bibfnamefont{V.}~\bibnamefont{Siruguri}},
  \bibnamefont{et~al.}, \bibinfo{journal}{arXiv:1707.09304}
  (\bibinfo{year}{2017}).

\bibitem[{\citenamefont{Fuchs et~al.}(2017)\citenamefont{Fuchs, Dey,
  Aslan-Cansever, Alfonsov, Kataev, Wurmehl, and B\"uchner}}]{Fuchs2017}
\bibinfo{author}{\bibfnamefont{S.}~\bibnamefont{Fuchs}},
  \bibinfo{author}{\bibfnamefont{T.}~\bibnamefont{Dey}},
  \bibinfo{author}{\bibfnamefont{G.}~\bibnamefont{Aslan-Cansever}},
  \bibinfo{author}{\bibfnamefont{A.}~\bibnamefont{Alfonsov}},
  \bibinfo{author}{\bibfnamefont{V.}~\bibnamefont{Kataev}},
  \bibinfo{author}{\bibfnamefont{S.}~\bibnamefont{Wurmehl}}, \bibnamefont{and}
  \bibinfo{author}{\bibfnamefont{B.}~\bibnamefont{B\"uchner}}
  (\bibinfo{year}{2017}).

\bibitem[{\citenamefont{Campbell et~al.}(1994)\citenamefont{Campbell, Amato,
  Gygax, Herlach, Schenck, Cywinski, and Kilcoyne}}]{Campbell1994}
\bibinfo{author}{\bibfnamefont{I.~A.} \bibnamefont{Campbell}},
  \bibinfo{author}{\bibfnamefont{A.}~\bibnamefont{Amato}},
  \bibinfo{author}{\bibfnamefont{F.~N.} \bibnamefont{Gygax}},
  \bibinfo{author}{\bibfnamefont{D.}~\bibnamefont{Herlach}},
  \bibinfo{author}{\bibfnamefont{A.}~\bibnamefont{Schenck}},
  \bibinfo{author}{\bibfnamefont{R.}~\bibnamefont{Cywinski}}, \bibnamefont{and}
  \bibinfo{author}{\bibfnamefont{S.~H.} \bibnamefont{Kilcoyne}},
  \bibinfo{journal}{Phys. Rev. Lett.} \textbf{\bibinfo{volume}{72}},
  \bibinfo{pages}{1291} (\bibinfo{year}{1994}).

\bibitem[{\citenamefont{Chen et~al.}(2017)\citenamefont{Chen, Svoboda, Zheng,
  Sales, Mandrus, Zhou, Zhou, McComb, Randeria, Trivedi et~al.}}]{Chen2017}
\bibinfo{author}{\bibfnamefont{Q.}~\bibnamefont{Chen}},
  \bibinfo{author}{\bibfnamefont{C.}~\bibnamefont{Svoboda}},
  \bibinfo{author}{\bibfnamefont{Q.}~\bibnamefont{Zheng}},
  \bibinfo{author}{\bibfnamefont{B.~C.} \bibnamefont{Sales}},
  \bibinfo{author}{\bibfnamefont{D.~G.} \bibnamefont{Mandrus}},
  \bibinfo{author}{\bibfnamefont{H.~D.} \bibnamefont{Zhou}},
  \bibinfo{author}{\bibfnamefont{J.~S.} \bibnamefont{Zhou}},
  \bibinfo{author}{\bibfnamefont{D.}~\bibnamefont{McComb}},
  \bibinfo{author}{\bibfnamefont{M.}~\bibnamefont{Randeria}},
  \bibinfo{author}{\bibfnamefont{N.}~\bibnamefont{Trivedi}},
  \bibnamefont{et~al.}, \bibinfo{journal}{arXiv:1707.06980}
  (\bibinfo{year}{2017}).

\end{thebibliography}

\end{document}